\documentclass[a4paper,10pt]{article}
\begin{document}
\title{Sliding Mechanism for Actin Myosin System}
\author{Bao-quan Ai, Xian-ju Wang,Guo-tao Liu and Liang-gang Liu\\
\sl \small  Department of Physics, ZhongShan University, GuangZhou, P. R. China\\
M. Nakano\\
\sl \small Department of Information Science, University of Occupational and Environmental Health, Japan\\
H. Matsuura\\
 \sl \small Department of Project Center, National Graduate Institute for Policy Studies, Japan\\}
\date{}
\maketitle

\begin{abstract}

Based on the Stochastic Inclined Rods Model (SIRM) proposed by H. Matsuura and M. Nakano, we study the microscopic motion of actin myosin system including the motion of the G-actin. Our model is composed of an inclined spring (rod), a myosin head, a myosin filament and G-actins. We discuss the stochastic resonance between the myosin head and random noise. The results of calculation show that the model can convert the random motion to one directional motion, and the myosin head works as a resonator of random noise which absorbs the energy through the stochastic resonance. The intermolecular potential between the myosin head and G-actin and the inclined rod play a key role for the muscle's motion. The energy consumed by the motor is directly supplied from the surroundings (i.e., the thermal motions of water molecules). 

\end{abstract}

\section{Introduction}
Biological systems are among the most challenging subjects for theoretical physicists as well as experimentalists or simulationists. More and more biologists and theoretical biophysicists have focused on studying the mechanism of living systems and mincing the biological structure to invent subtle artificial instruments. Nowadays it is exciting to study a problem how motor protein gets a energy of movement and what makes myosin slide in one direction along the filaments. Based on different system, three different families of motor proteins have been identified \cite{1}\cite{2}\cite{3}: Kinesis and dyneins move along tubulin filaments \cite{4},  myosin moves along actin filaments \cite{5}\cite{6}. The motion mechanism of these motors can be described in general as follows \cite{1}\cite{7}\cite{8}: firstly the motor protein binds Adenosinetriphoshpate (ATP), then hydrolyzes the bound ATP and absorbs the energy, subsequently, it releases the products Adenosinediphosphate (ADP) and Phosphate (P). So the motors move constantly within chemical cycle. The molecular motors play a key role in transducing chemical energy into mechanical work at a molecular scale. \\
 \indent In 1954 H. F. Huxley, H. E. Huxley and R. M. Simmons \cite{9}\cite{10}\cite{11} proposed the rotation cross bridge model to study muscular mobility. In these earliest theoretical descriptions, the asymmetry of the system was introduced via asymmetric transition rates and specific conformational changes were also discussed. This model has been widely spread in past two decades and has become a standard model even in college textbooks, but extensive experimentations have failed to find evidence for a large conformational change in myosin head during the force generation process \cite{12}. So it is worth while to reconsider an old idea,  proposed by A. F. Huxley in 1957,  the motor protein uses thermal energy that exists at physiological temperature to perform work \cite{13},  such as the thermal ratchet model \cite{14} \cite{15}, the isothermal ratchet model \cite{16}.\\
\indent Recently, there have been many reports that discuss the relation between the muscle's movement and Stochastic Resonance (SR) or thermal noise \cite{17}\cite{18} \cite{19}\cite{20}. These explanations, however, somewhat abstract from the viewpoint of biology, and these models are lacking in dynamic descriptions of actin myosin interaction.\\  
    \indent The newly resonance models for muscle's motion was named the Stochastic Inclined Rods Model (SIRM) proposed by H. Matsuura and M. Nakano \cite{21}\cite{22}\cite{23}\cite{24}. The energy of the motion was supplied from the random noise and the system always moves to one direction by using of stochastic resonance. The movement of the system does not break the second law of the thermodynamics, because the actin myosin system is open to the surroundings and the energy flows in from their surroundings. The SIRM presents a perfectly sliding mechanism for the actin myosin system.\\
\indent The main aim of our paper is to study the motion of actin myosin system including the G-actin's motion. We discuss the stochastic resonance of the myosin head and give the numerical results for the motion of the system. We find that there is a relative sliding motion between the G-actin and the myosin owing to the intermolecular potential and the structure of the inclined rod. \\

\section{Model and Formalism}

 The structure of the actin myosin system is shown in Fig. 1\cite{21} together with structure of the muscle. It consists of the chain structure of the actin of length around 100 $nm$ and a bundle of hundreds of myosin which has an inclined rod and a head of 0.45 million Dalton. The whole system exists in a water solution and the interaction with water molecules cannot be neglected.\\ 
\begin{center}
\begin{tabular}{|c|}\hline
Fig. 1\\ \hline
\end{tabular}
\end{center}

\indent In order to explain the contraction of the muscle,  namely solve the motion of actin myosin system,  we simulate it by a mechanical model as shown in Fig. 2. We call this model as stochastic inclined rod model (SIRM).  The motion mechanism of the SIRM is as follows: firstly, ATP is hydrolyzed and releases the energy which makes the random noise interacts with the myosin head and myosin head obtains the energy from random noise then the head vibrates, it collides with a G-actin and obliquely kicks the G-actin sphere, because the direction of vibration is inclined against the line of the actin fibers and myosin molecules obtain the propellant force along the direction of the fibers. In this way, the filament can move to one direction while the G-actin moves to opposite direction.  The actin myosin system can move to one direction owing to the structure of inclined rods of myosin and the intermolecular potential between the myosin head and the G-acin.\\
\begin{center}
\begin{tabular}{|c|}\hline
Fig. 2\\\hline
\end{tabular}
\end{center}
\indent 
According  to Fig. 2 we can construct the equations of the motion for the head of myosin. 
\begin{equation}
\label{e1}
m{\frac{\partial^{2} x}{\partial t^{2}}=-{\frac{\partial (U_{a}+U_{s})}
{\partial x}}+F_{x}(t)-\alpha{\frac{\partial x}{\partial t}}}.
\end{equation}
\begin{equation}
\label{e2}
m{\frac{\partial^{2} y}{\partial t^{2}}=-{\frac{\partial (U_{a}+U_{s})}
{\partial y}}+F_{y}(t)-\beta{\frac{\partial y}{\partial t}}}.
\end{equation}
\indent With regard to the center of gravity $(x_{2}, y_{2})$ of the filament, we set a similar equation:
\begin{equation}
\label{e3}
M{\frac{\partial^{2} x_{2}}{\partial t^{2}}=-{\frac{\partial U_{s}}
{\partial x_{2}}}+F_{x_{2}}(t)-\eta{\frac{\partial x_{2}}{\partial t}}}.
\end{equation}
The variable $y_{2}$ is fixed since the myosin filament does not significantly move along the $y$-direction compared to the $x$-direction.\\
\indent The equation of the G-actin is shown as follow:\\
\begin{equation}
\label{e4}
M_{g}{\frac{\partial^{2} x_{g}}{\partial t^{2}}=-{\frac{\partial U_{a}}
{\partial x_{g}}}+F_{x_{g}}(t)-\gamma{\frac{\partial x_{g}}{\partial t}}}.
\end{equation}\\
\indent Where $F_{x}(t), F_{y}(t), F_{x_{2}}(t),F_{x_{g}}(t) $ are fluctuation of the thermal noise, $m, M, M_{g}$ are the masses of myosin head,  myosin filament and the G-actin, respectively, $\alpha, \beta, \eta, \gamma$ are viscous constants, and the fluctuation forces follow fluctuation-dissipative relation \cite{25}\cite{26}\cite{27}:\\
\begin{equation}
\label{e5}
<F_{a}(t)>=0.
\end{equation}
\begin{equation}
\label{e6}
<F_{a}(t)F_{b}(s)>=2k_{B}T{\zeta}{\delta}_{a,b}{\delta}(t-s).
\end{equation}
where $a,b=x, y, x_{2}, x_{g}$; ${\zeta}={\alpha}, {\beta}, {\eta}, {\gamma}$;  $k_{B}$ is Boltzmann constant.  $T$ is absolute temperature. $t, s$ are time.\\
\indent As for the intermolecular potential between the myosin head and G-acitn we adopt Jennard-Jone potential:\\
\begin{equation}
\label{e7}
U_a={\sum\limits_{i=0}^{i=n}(pr_{i}^{-12}-qr_{i}^{-6})U}.
\end{equation}
\indent Where $r_{i}=\sqrt{(x-x_{i})^{2}+(y-y_{i})^{2}}-R$; $x_{i}=x_{g}+iR$; $y_{i}=y_{g}$;   $(x_{g},y_{g})$ stands the coordinates of the center of the G-actin and R is its radius. 
 $p, q, U$ are the parameters of the potential.\\
\indent The potential of the myosin rod is approximately expressed as follows:\\
\begin{eqnarray}
\label{e8}
U_s=K_{l}\exp(-\sqrt{(x-x_{2})^{2}+(y-y_{2})^{2}}+L)\nonumber\\
\nonumber  +K_{l}\exp(\sqrt{(x-x_{2})^{2}+(y-y_{2})^{2}}-L)\nonumber\\
 +{1\over 2}K_{\theta}(\theta-\theta_{0})^{2}.
\end{eqnarray}
\indent Where $K_{\theta},K_{l}$ are parameters of potential of the spring 
 and $L$ is  nature length of the spring,  ${\tan(\theta)}={x_{2}-x\over y_{2}-y}, \theta,{\theta}_{0}$ are current angle and initial angle between the rod and level axis. $ (x_{2},y_{2})$ is the center of gravity of myosin bundle.\\
\section{stochastic resonance}

We discuss the stochastic resonance between the myosin head and the thermal noise to show that SIRM moves to one direction with accepting energy from thermal noise. Because the above equations are so complex and highly nonlinear that we cannot obtain a strict analytical solution, we only discuss the motion of the myosin head and set up the following simple equation for the head, where the  $j$-th G-actin is the nearest one to the head:\\
\begin{equation}
\label{e9}
m{\frac{\partial^{2} x}{\partial t^{2}}=12pU(x-x_{j})^{-13}-6qU(x-x_{j})^{-7}-2(x-x_{2}+L)K+F_{x}(t)-\alpha{\frac{\partial x}{\partial t}}}.
\end{equation}
\indent Because the fluctuation $F(t)$ can be given in the following form \cite{22}:\\

\begin{equation}
\label{e9}
F_{a}(t)={\sqrt{k_{B}T\zeta\over\pi}}{\sum\limits_{j}\rho_{a}^{j}\exp(i\Omega_{j} t)}.
\end{equation}
\begin{equation}
\label{e9}
<F_{a}(t)F_{b}(s)>={k_{B}T\zeta\over\pi}{\sum\limits_{j,l}\delta_{a,b}\delta^{j,l}\exp i(\Omega ^{l} s-\Omega^{j} t)}.
\end{equation}
\begin{equation}
\label{e12}
<{\rho_{a}^{j}}^{*}\rho_{b}^{l}>=\delta_{a,b},     <{\rho_{a}^{j}}^{*}>=<\rho_{b}^{l}>=0.
\end{equation}
\indent If we assume that the thermal noise contains all of frequency, one of the noise frequency is expressed as follows:\\
\begin{equation}
\label{e13}
{F_{x}(t)\over m}\rightarrow {f\over m }\exp (i\Omega t).
\end{equation}
\indent Thus Eq.(9) can be given:\\
\begin{equation}
\label{e14}
{\frac{\partial^{2} \xi}{\partial t^{2}}}+2\mu{\frac{\partial \xi}{\partial t}}+\omega^{2}\xi={f\over m}\exp(i\Omega t)-\varepsilon\alpha\xi^{2}-\varepsilon\beta\xi^{3}.
\end{equation}
\indent Where $\xi=x-x_{2}+L$, $\xi_{j}=x_{j}-x_{2}+L$, $2\mu={\alpha\over m}$.
\begin{equation}
\label{e15}
\omega_{0}^{2}={2K\over m}+{156pU\over m\xi_{j}^{14}}-{42qU\over m\xi_{j}^{8}}, \varepsilon\alpha={2184pU\over 2m\xi_{j}^{15}}-{336qU\over 2m\xi_{j}^{9}}, \varepsilon\beta={32760pU\over 3!m\xi_{j}^{16}}-{3024qU\over 3!m\xi_{j}^{10}}.
\end{equation}
\indent Here we use Taylor expansion for the intermolecular force of the G-actin and ignore the higher order terms.\\
\indent When $\varepsilon=0$ and $\mu^{2}>\omega_{0}^{2}$ we can get the solution of the Eq. (14) as follows:\\
\begin{equation}
\label{e16}
\xi=a_{0}e^{-\mu t}\cos(\omega t+\vartheta_{0})+ae^{i(\Omega t-\delta)}.
\end{equation}
\indent Where $\omega=\sqrt{\omega_{0}^{2}-\mu^{2}}$,  ${a}={f\over{m\sqrt{(\omega_{0}^{2}-\Omega^{2})^{2}+4\mu^{2}\Omega^{2}}}}$, ${\tan\delta}=-{2\mu\Omega\over{\omega_{0}^{2}-\Omega^{2}}}$, $a_{0},\vartheta_{0}$ are determined by initial conditions.\\
\indent If $\varepsilon\neq 0$, we can use a effective frequency for the Eq. (14)\cite{28}:\\
\begin{equation}
\label{e17}
\omega^{*}=\omega_{0}+{3\varepsilon a^{2}\over 8\omega_{0}^{2}}.
\end{equation}
\indent So we can rewrite the Eq.(14) as follow:\\
\begin{equation}
\label{e18}
{\frac{\partial^{2} \xi}{\partial t^{2}}}+2\mu{\frac{\partial \xi}{\partial t}}+{\omega^{*}}^{2}\xi={f\over m}\exp(i\Omega t).
\end{equation}
\indent So the solution of the Eq. (18) should have the same form as that of Eq.(14), but the amplitude $a$ should be satisfied the following relation:\\
\begin{equation}
\label{e19}
{a^{2}}={({f\over m})^{2}\over{({\omega^{*}}^{2}-\Omega^{2})^{2}+4\mu\Omega^{2}}}
\end{equation}
\indent If we assume the frequency $\Omega=\omega_{0}+\Delta$, we can obtain:\\
\begin{equation}
\label{e20}
({\omega^{*}}^{2}-\Omega^{2})^{2}=4\omega_{0}({3\varepsilon a^{2}\over 8\omega_{0}}-\Delta)^{2}.
\end{equation}
\indent We can rewrite the Eq. (19) as follow:\\
\begin{equation}
\label{e20}
\sigma[(\sigma-\Delta)^{2}+({\mu\over\varepsilon})]=F.
\end{equation}
\indent Where ${\sigma}={3a^{2}\over 8\omega_{0}}$, ${F}={3f^{2}\over 32m^{2}\omega_{0}^{3}\varepsilon^{2}}$. From the Eq. (21) we can get the relation between the resonance amplitude and the shift frequency from the resonance point.\\
\begin{center}
\begin{tabular}{|c|}\hline
Fig. 3\\\hline
\end{tabular}
\end{center}
\indent 
The figure shows the relation between the square of resonance amplitude $\sigma$ and the shift frequency from the resonance point $\Delta$,  which is the solution of the Eq. (21). If the amplitude of noise $F$ is small, the peak occurs at $\Delta=0$, as the amplitude of the noise increases, the shape of $\sigma$ changes gradually while it keeps the character that has the only one maximum, which truly shows the existence of the stochastic resonance between the myosin head and the thermal noise.\\
\indent After a sufficient time, the damping terms of the Eq.(16) go to zero, and only oscillational term $ae^{i(\Omega t-\delta)}$ survives. It is clear that these equations (Eq. (1)-Eq. (4)) do not have an energy source except the thermal noise term $F(t)$. In the case,  the energy of the myosin  has the constant value which is determined by the amplitude $a$. And the average absorbed energy per unit time will be $E(\Omega)=2\Omega^{2}\mu m a^{2}$.

\section{Numerical results}

We adopt a numerical method to solve the above equations. The parameters of the equations are shown in the Table.1.\\
\begin{center}
\begin{tabular}{|c|}\hline
Table. 1\\\hline
\end{tabular}
\end{center}
\indent The numerical results are shown in Fig. 4-Fig. 8.\\
\begin{center}
\begin{tabular}{|c|}\hline
Fig. 4\\\hline
\end{tabular}
\end{center}
\indent The figure shows the horizontal position of myosin head, the trace has irregularity and randomness like Brownian particles, but movement is a translacational one as a whole.\\
\begin{center}
\begin{tabular}{|c|}\hline
Fig 5\\\hline
\end{tabular}
\end{center}
\indent The solid line shows the vertical motion of the myosin head and the stretching motion of the rod is described with the dash line. From the figure we can know that the stretching vibration of the spring has irregularity and the collision between the G-actin and the myosin head distorts their relative motion from a trigonometric function.\\
\begin{center}
\begin{tabular}{|c|}\hline
Fig 6\\\hline
\end{tabular}
\end{center}
\indent 
The dash line shows the motion of the filament and the solid line shows the position of the G-actin. From the figure we can know there is a relative sliding motion between the filament and the G-actin. The filament can move to one direction while the G-actin moves to opposite direction.\\
\begin{center}
\begin{tabular}{|c|}\hline
Fig 7\\\hline
\end{tabular}
\end{center}
\indent The figure gives the relative between the horizontal movement of myosin filament and the initial angle. From the figure we can know that the motor can move fastest when the initial angle is nearly equal to $45^{\circ}$ . But if the initial angle is too small or too big, the motor moves slowly, even cannot move. So the initial angle of the inclined rod is important for SIRM to move to one direction.\\

\begin{center}
\begin{tabular}{|c|}\hline
Fig 8\\\hline
\end{tabular}
\end{center}
\indent From the figure we can know that the shape of the G-actin is important for the system to move. If the G-actin if flat ($R=0$) the SIRM can not move, on the other hand, if the radius of the G-actin is too big ($R=40$). The system can  not move, either. The motor can move at the highest speed at $R=20$.\\

\section{Summary and conclusion}

Based on SIRM we present a sliding mechanism for the actin myosin system and set up the dynamic equations for the model. Because equations (Eq. (1) - Eq. (8) ) are so complex and highly nonlinear that we cannot obtain a strict analytical solution, we adopt a numerical method to solve these two dimensional Langevin equations.  We discuss the motion of the myosin head and find there is a stochastic resonance between the head and the noise. Our model propose a microscopic mechanism for the actin myosin system: firstly, the thermal noise interacts with the myosin head, and the resonance occurs between the myosin head and the noise, then it collides with a G-actin and obliquely kicks the G-actin sphere, because the direction of vibration is inclined against the line of the actin fibers, myosin molecules obtain the propellant force along the direction of the fiber and the myosin can move to one direction.\\
\indent From the numerical results we can know there is a relative sliding motion between the filament and the G-actin. The filament can move to one direction while the G-actin moves to opposite direction. When the system absorbs the energy from the thermal noise constantly through stochastic resonance, the intermolecular potential and the inclined rod make the filament move to one direction. The system can convert the random noise to one directional motion.  SIRM is thermally open to the outer surroundings and it has a outer heat source of ATP.  The heat consumed by SIRM is directly supplied from the surroundings and the energy provided to the surroundings comes from the outer source or reservior through a general energetic flow. From a macroscopic point of view, SIRM can move by getting heat from its surroundings.\\
\indent
Noise in dynamical systems is usually considered a nuisance. But in certain nonlinear systems including electronic circuits and biological sensory apparatus the presence of noise can in fact enhance the detection of the weak signals. From the above results, we point out the possibility that the actin myosin system can use thermal noise around for the movement in similar way. This viewpoint is completely new as the traditional viewpoint thinks the water disturbs the motion as viscosity.\\

\newpage
\section{Figure captions}
\baselineskip 0.3in
{\bf{Table. 1.}} The parameters of the equations. $\alpha, \beta, \eta,\gamma$ are viscous constants, $M, m, M_{g}$ are the mass of myosin bundle and myosin head,  respectively, $L$ is nature length of the spring,  $p, q, U$ are the parameters of the potential,  $K_{\theta},K_{l}$ are tangent constant and radial constant of the spring. \\
{\bf{Fig. 1.}} Actin myosin system as an element of the muscle shown in the bottom of the figure. The structure of muscle is also shown from the upper level to lower level (see ref.[21]).\\
{\bf{Fig. 2.}} The model for actin myosin system. The coordinates of the head of myosin and the end of myosin connected to the filament are $(x, y)$ and $(x_{2}, y_{2})$, respectively. The body of the myosin is assumed to be represented as  a spring.\\
{\bf{Fig. 3.}} The figure shows the solution of Eq. (21) and give the relation between the square of resonance amplitude ${\sigma}={3a^{2}\over 8\omega_{0}}$ and the shift frequency $\Delta$ from resonance point.  From the up to the bottom, different lines correspond to ${F}={3f^{2}\over 32m^{2}\omega_{0}^{3}\varepsilon^{2}}$: 20, 15, 10, 5, 1, respectively.\\  
{\bf{Fig. 4.}} The figure shows the horizontal motion of the myosin head (see Eq. (1)).\\
{\bf{Fig. 5.}} The solid line shows the vertical movement of the myosin head and the dash line presents the stretching displacement $s$ of the rod and $s=\sqrt{(x-x_{2})^{2}+(y-y_{2})^{2}}-L$ (see Eq. (2)).\\
{\bf{Fig. 6.}} The dash line gives the movement of the filament ($x_{2}$) and the solid line shows the position ($x_{g}$) of the G-actin (see Eq. (3) and Eq. (4)).\\
{\bf{Fig. 7.}} The curve shows the horizontal displacement of the myosin filament at different initial angle $\theta_{0}$ ($t=60 unit time$).\\
{\bf{Fig. 8.}} The curves give the relation between the radius of G-actin and the displacement of filament in given time, different lines correspond to different radius $R$ shown in the figure.\\
\begin {center}
Table. 1. The Parameters of the Equations.\\
\end{center}

\tabcolsep 0.15in
\begin{tabular}{|c|c|c|c|c|c|}\hline
1 unit time &$10^{-11}$s& $\alpha, \beta, \eta, \gamma$ &1& $U$&20000\\\hline
1 unit length &  $10^{-9}$m &  $ p, q$&  100& $m$ & 1000 \\\hline
 1 unit mass& $2.5\times10^{-21}$kg &$L$ &100& $M$&3000 \\\hline
$K_{\theta}$ & 500000& $K_{l}$ & 500 &$M_{g}$ &10000\\\hline
\end{tabular}

\newpage

\begin{thebibliography}{99}
\baselineskip 0.2in
\bibitem{1}F. Julicher, A. Ajdari and J. Prost, Rev. Mod. Phys., 69 (1997) 1269.
\bibitem{2}T. Kreis and R. Vale, Cytoskeletal and motor proteins, Oxford University Press (1993).
\bibitem{3}K. Y. Wang, S. B. Xue, H. T. Liu, Cell Biology, (Beijing Normal University Press, in Chinese ), (1998) 238-246.
\bibitem{4}E. Mandelkow and A. Hoenger, Current Opinion in Cell Biology, 11 (1999) 34.
\bibitem{5}G. G. Borisy and T. M. Svitkina, Current Opinion in Cell Biology, 12 (2000) 104.
\bibitem{6}T. Yanagida, K. Kitamura,H. Tanaka, A. Hikikoshi lwane and S. Esaki, Current Opinion in Cell Biology, 12 (2000) 20.
\bibitem{7}T. L. Hill, Prog. Biophys. Mol., 28 (1994) 267.
\bibitem{8}J. A. Spudich, Nature, (1990) 284.
\bibitem{9}A. E. Huxley and R. M. Simmons, Nature, 233 (1971) 533.
\bibitem{10}A. F. Huxley and R. Niedergerke, Nature, 173 (1954) 971.
\bibitem{11}A. F. Huxley and J. Hanson, Nature, 173 (1954) 973.
\bibitem{12}T. Yanagida and E. Homsher, Plenum Press, New York, (1984) 833.
\bibitem{13}A. F. Huxley, Prog. BioPhys., 7 (1957) 255.
\bibitem{14}M. O. Magasco, Phys. Rev. Lett., 71 (1993) 1477.
\bibitem{15}R. D. Astumian, Phys. Rev. Lett., 72 (1994) 1766.
\bibitem{16}A. Parmeggiani, J. Frank, A. Ajdari and J. Prost, Phys. Rev. E, 60 (1999) 2127
\bibitem{17}W. Kurt and M. Frank, Nature, 373 (1995) 33-36.
\bibitem{18}J. J. Collins, C. C. Chow and T. T. Imhaff, Nature, 383 (1996) 770.
\bibitem{19}J. Rousslet, L. Solomw, A. Ajdoari and J. Prost, Nature, 370 (1994) 446-449.
\bibitem{20}G. Hu, A. Daffertsshofer, H. Haken, Phys. Rev. Lett., 76 (1996) 4874-4877.
\bibitem{21}H. Matsuure and M. Nakano, Biomedical and Human Science, 3 (1) (1997) 47.
\bibitem{22}H. Matsuure and M. Nakano, Information, Vol.3(2) (2000) 203-230.
\bibitem{23}H. Matsuure and M. Nakano, Proc. of the IEEE-EMBS Asia-Pacific Conference on Biomedical Engineering, (2000) 377.
\bibitem{24}H. Matsuura and M. Nakano, AIP(American Institute of Physics) Conference
         Proceeding 519, Statistical Physics (2000) 557-559.
\bibitem{25}G. E. Uhelenbeck and Ornsteinn, Phys. Rev., 36 (1930) 823.
\bibitem{26}M. C. Wang and G. E. Uhelenbeck, Rev. Mod. Phys., 17 (1954) 323.
\bibitem{27}K. Tawada and K. J. Sekimoto, Theor. Biol., 150 (1991) 193. 
\bibitem{28}L. K. Liu, S. D. Liu, Nonlinear Equations in Physics, (2000) 107.
\end{thebibliography}
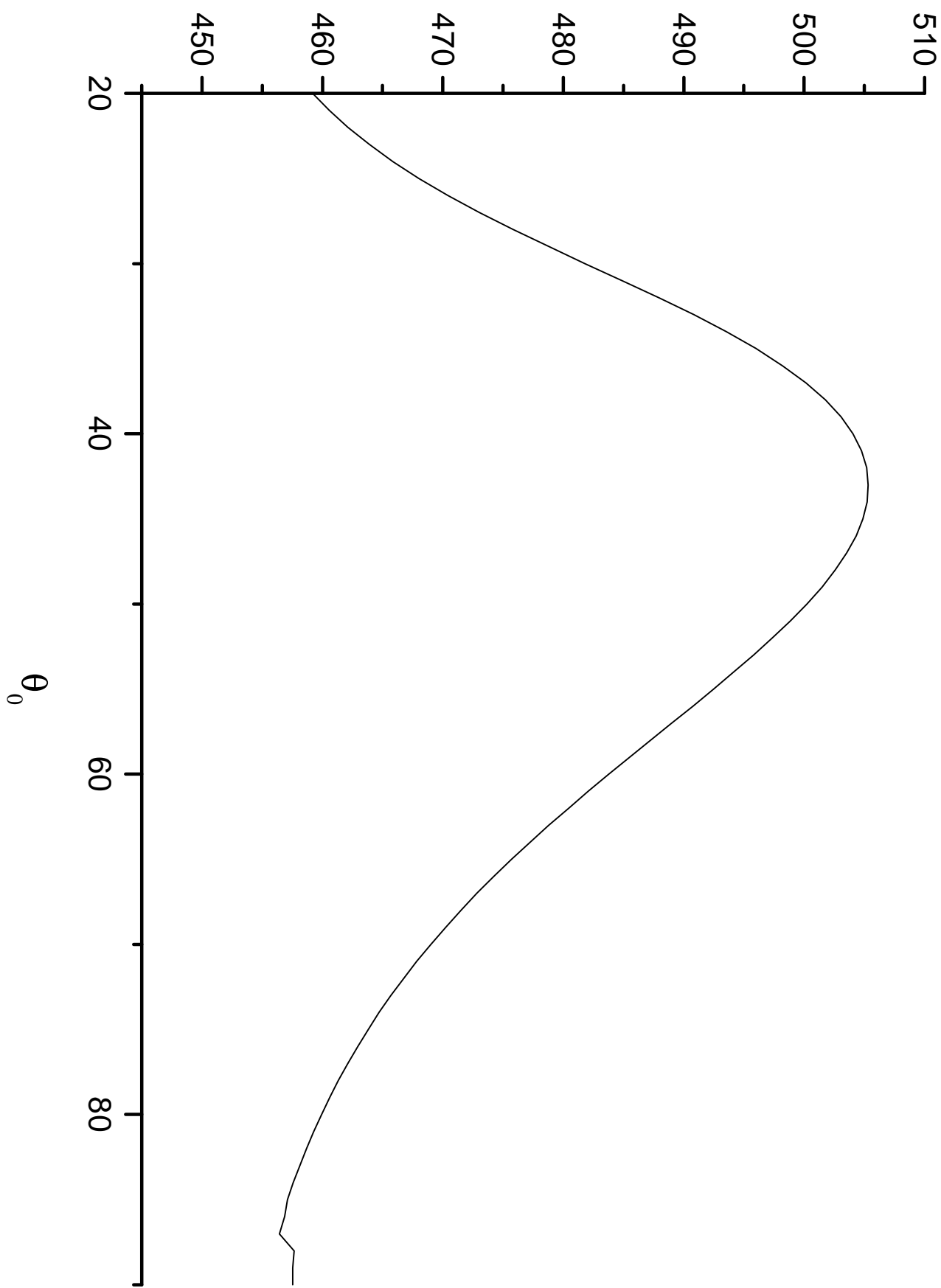
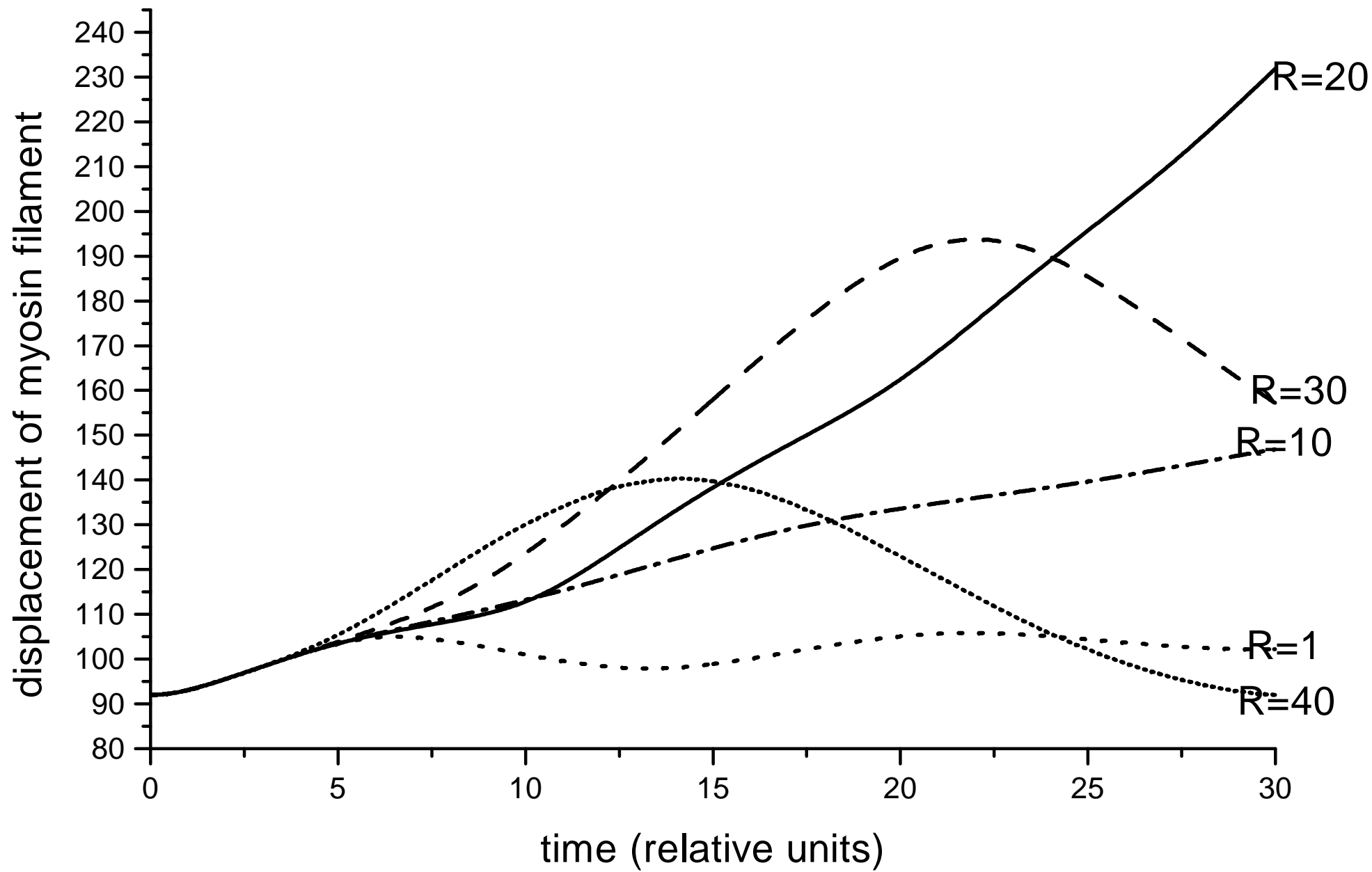
\end{document}